\documentclass[12pt]{article}
\usepackage{graphicx,amssymb,epsfig}

\makeatletter
\renewcommand\section{\@startsection {section}{1}{\z@}%
                                 {-3.5ex \@plus -1ex \@minus -.2ex}%nn
                                   {2.3ex \@plus.2ex}%
                                   {\normalfont\large\bfseries}}
\renewcommand\subsection{\@startsection{subsection}{2}{\z@}%
                                   {-3.25ex\@plus -1ex \@minus -.2ex}%
                                     {1.5ex \@plus .2ex}%
                                     {\normalfont\bfseries}}
\renewcommand\subsubsection{\@startsection{subsubsection}{3}{\z@}%
                                   {-3.25ex\@plus -1ex \@minus -.2ex}%
                                     {1.5ex \@plus .2ex}%
                                     {\normalfont\itshape}}
\makeatother

\newcounter{multieqs}

\newcommand{\bq}{\begin{equation}}
\newcommand{\fq}{\end{equation}}
\newcommand{\bqr}{\begin{eqnarray}}
\newcommand{\fqr}{\end{eqnarray}}

\begin{document}
\thispagestyle{empty}
\begin{flushright}
\parbox[t]{2in}{CERN-TH/2002-359\\
MAD-TH-02-3}
\end{flushright}

\vspace*{0.5in}

\begin{center}
{\Large \sc (Re)Constructing Dimensions}

\vspace*{0.5in} 
{\large Ra\'ul Rabad\'an${}^{1}$
and Gary Shiu${}^{2}$}\\[.3in]
{\em ${}^1$ Theory Division CERN, CH-1211 Gen\`eve 23, Switzerland\\[.1in]
${}^2$ Department of Physics,
  University of Wisconsin,
Madison, WI 53706, USA}\\[0.5in]
\end{center}

\begin{center}
{\bf
Abstract}
\end{center}
Compactifying a higher-dimensional theory defined in $R^{1,3+n}$ 
on an n-dimensional manifold ${\cal M}$ results in a spectrum of 
four-dimensional (bosonic) fields
with  masses 
$m^2_i = \lambda_i$, where $- \lambda_i$ are the eigenvalues of the 
Laplacian on the compact manifold. The question we 
address in this paper
is the inverse: given the masses of the Kaluza-Klein fields in four dimensions, what can we say about the 
size and shape (i.e. the topology and the metric) of the compact manifold?
We present some examples of isospectral manifolds (i.e., different manifolds which give rise
to the same Kaluza-Klein mass spectrum). 
Some of these examples are Ricci-flat, complex and K\"{a}hler and so
they are isospectral backgrounds for string theory.
Utilizing results from finite spectral geometry, we also discuss the accuracy
of reconstructing the properties of the compact manifold (e.g., its dimension, volume, and
curvature etc) from measuring
the masses of only a finite number
of Kaluza-Klein modes. 

\vfill

\hrulefill\hspace*{4in}

{\footnotesize
Email addresses: Raul.Rabadan@cern.ch, shiu@physics.wisc.edu.}

\newpage

\tableofcontents

\section{Introduction}

In recent years, motivated by the brane world scenario \cite{HW,ADD,ST,Lykken,Ovrut,RS}, 
there has been an enormous interest
in exploring the experimental signatures of extra dimensions \cite{Hewett}.
In the context of string theory, information about
the size and shape of the extra dimensions
could shed light on the nature of our string vacuum.
Indeed, 
one of the outstanding challenges in connecting string theory with low energy physics
is 
to understand how the enormous degeneracy of string vacua is lifted. 
 Although the underlying mechanism which selects our string vacuum is unknown at present,
over the years we have gained 
a great deal of insights into the physical
implications of string theory
by exploring
compactifications of string and M theory on various manifolds.
Since the spectrum of 
particles
and their interactions in the low energy theory 
are determined by the size and shape of the extra dimensions,
we can use phenomenological constraints to focus our attention on the subset of
manifolds that are relevant in describing our observed four-dimensional physics.
For example, in compactifying the weakly coupled heterotic string on a Calabi-Yau manifold with
the standard embedding, one finds that 
the number of
generations of chiral matter is determined by the Euler character  whereas
the Yukawa couplings are given by the intersection numbers of cycles in the 
Calabi-Yau space. 
In brane world models 
where chiral fermions arise from
intersecting branes \cite{CSU1,CSU2,CSU3,Berlin1,Madrid1,Madrid2,Berlin2}, 
the number of chiral generations is again determined by a topological
quantity, namely, the intersection number between
different D-branes \footnote{In a T-dual picture, these numbers are related to Dirac indexes \cite{fluxes}, that again are topological quantities.}.
The basic premise of string phenomenology is to make use of
some geometrical/topological insights to construct string models
that satisfy some basic physical 
requirements (e.g., models with three generations, 
the correct structure of Yukawa couplings and gauge couplings, etc) from which we can
extract their low energy predictions.
In addition to the Standard Model-like spectrum of light fields, different compactifications 
generically give rise to different spectra of Kaluza-Klein (KK) modes, winding modes as
well as stringy states.
Therefore, if we were able to measure the mass spectrum of
these heavy modes (i.e., the size and shape of the extra dimensions), 
we could further
constrain the types of compactifications.
This is also the philosophy behind the studies of the experimental signatures of
different extra dimension scenarios
as the KK spectra are different for 
large extra dimensions \cite{ADD},
warped compactifications \cite{RS}, compact hyperbolic manifolds \cite{hyper},
and etc.

The question we address in this paper is the reverse, namely, {\it given 
the mass spectrum of the 
Kaluza-Klein (KK) modes, how well can we reconstruct the 
extra 
dimensions?} As we shall see, there exist 
distinct manifolds which nonetheless have the same set of eigenvalues of the Laplacian.
Therefore, compactifications on these {\it isospectral manifolds} result in
identical KK mass spectrum.
Interestingly,
some of the isospectral manifolds are Ricci-flat, complex, and 
K\"{a}hler.
Hence
they provide isospectral backgrounds not only for point particles but also 
for the propagation of strings.
In reality, we cannot measure the full KK spectrum and so
perhaps a physically more relevant question is:
how much information about the properties of a compact manifold 
(e.g., its dimension, volume, and curvature etc) can we extract from measuring the masses of a finite number of
Kaluza-Klein modes?
Utilizing results from the subject of finite spectral geometry, we  obtain some quantitative
estimate of the accuracy in reconstructing the geometrical properties (such as the
dimension, volume and curvature) of the 
compact manifold.

The fact that geometrical interpretations are not unique is not a new surprise in string theory
-- one of the notable examples of such non-uniqueness is mirror symmetry.
It has long been conjectured \cite{Dixon,LercheVafaWarner} and demonstrated by explicit examples \cite{GreenePlesser,Candelas}
that in string theory there are topologically distinct manifolds which give rise
to identical physical models.
The equivalence of mirror manifolds is due to the extended nature of strings --
roughly speaking, the KK modes and the winding modes are interchanged
under mirror symmetry.
Compactifications on isospectral manifolds
correspond to a simpler form of equivalence -- 
the mass spectra of the KK modes and the winding modes of 
isospectral manifolds are identical individually. Therefore such equivalence holds even
in the point particle ($\alpha' \rightarrow 0$) limit.
However, we note that unlike mirror manifolds which give rise to physical models with
identical spectrum as well as interactions, compactifications on 
isospectral manifolds could differ at the
level of interactions and so these manifolds can be disentangled with additional
measurements such as  the couplings between different KK modes.

An important lesson drawn from the idea of deconstruction \cite{deconstruct1,deconstruct2} \footnote{Some similar ideas were considered in \cite{Halpern}.}
is that from the mass spectrum of only a finite number of KK 
modes (the low lying modes), we cannot distinguish between
a higher-dimensional theory and a four-dimensional gauge theory which becomes
strongly coupled in the infrared.
Here, we shall see that even the full KK mass spectrum
does not give us sufficient information to reconstruct the compact manifold.

\vspace{0.5cm}
This paper is organized as follows: in section \ref{scalar} we revisit the
general properties of a free scalar in a compact space. This will be our toy model to discuss some general properties of compactifications. 
Then, in section \ref{heat}, we analyze the asymptotic expansion of the heat kernel given properties of the manifold that we can read from the spectrum of the Laplacian. In section \ref{examples}, we compute the heat kernel for some simple examples: flat tori and spheres. In particular, we will see how to read off
different information about the manifold from these functions. In section \ref{forms}, we will 
discuss how the Laplacian acting on forms can give more information. Then we will describe some properties of different manifolds with the same spectrum, i.e. isospectral manifolds. In section \ref{interactions} we will briefly describe how interactions can distinguish between isospectral manifolds. We then discuss the implications of isospectral manifolds to string theory: we will consider some examples of isospectral manifolds where the conformal theory can be constructed, e.g. flat tori. Then we will discuss some properties of general string backgrounds, in particular, how strings will not be able classically to distinguish isospectral manifolds from their spectrum. Finally, in section \ref{finite}, we will address the question of what can we say with only a finite number of eigenvalues, i.e., the 
low-lying Kaluza-Klein masses.

\section{Scalar Field on a Compact Manifold}
\label{scalar}

Let us take the simplest example: consider a free scalar field living on 
$R^{1,3} \times {\cal M}$, where we take ${\cal M}$ to be a compact manifold 
without boundary of dimension $n$:

\begin{equation}
\phi :  R^{1,3} \times {\cal M}  \longrightarrow R
\end{equation}

From the four dimensional perspective, we get a set of scalar fields with a 
mass$^2$ $ m^2_i = \lambda_i$, where $-\lambda_i$ are the eigenvalues of the Laplacian 
operator on the compact  Riemannian manifold, i.e. a manifold plus a metric 
on it $({\cal M}, g)$:

\begin{equation}
\Delta_g \phi_i = -\lambda_i \phi_i.
\end{equation} 

The eigenfunctions, $\phi_i$, can be taken to be an orthonormal basis for the harmonic functions on ${\cal M}$:

\begin{equation}
\int_{\cal M} \phi_i \phi_j = \delta_{ij}.
\end{equation}

\vspace{0.5cm}

The spectrum of the Laplacian operator, i.e. the set of its eigenvalues counted with the corresponding multiplicities $\{ \lambda_i \}$, is a discrete sequence of real positive number with  the following properties:
\begin{itemize}
\item There is a unique zero mode which is just a constant 
$\phi_0 = 1/\sqrt{Vol({\cal M})}$,
\item The eigenvalues form a numerable, infinite set: 
$\lambda_i \leq \lambda_{i+1}$,
\item The limit $\lim_{i \rightarrow \infty} \lambda_i = \infty$,
\item The integral of the eigenfunctions in the compact space vanishes except for the zero mode:
\begin{equation}
\int_{\cal M} \phi_i = \sqrt{Vol({\cal M})} \delta_{i0}.
\end{equation}
\item In the limit where $k \rightarrow \infty$ the eigenvalues $\lambda_k$ satisfy the Weyl asymptotic formula:
\begin{equation}
\lim_{k \rightarrow \infty} \frac{\lambda_k}{k^{2/n}} = \frac{c_n}{(Vol ({\cal M}))^{2/n}}
\end{equation}
where $c_n = \frac{(2 \pi)^2}{\omega_n^{2/n}}$ and $\omega_n$ is the volume of the unit ball in $R^n$. Then at very high masses, $k \rightarrow \infty$, the variation of the number of eigenvalues with respect to the variation of the masses is:
\begin{equation}
\delta N = \frac{n Vol ({\cal M})}{c_n^{n/2}} m^{n-1} \delta m 
\end{equation}
as expected. If the number of dimensions is bigger  the number of modes grows very quickly.

\end{itemize}   

Now, the questions that we would like to consider is the following: given the 
spectrum, what information about the compact manifold can we reproduce? 
More specifically,
what are the geometric properties that are determined by the spectrum? Are 
there two or more Riemannian manifolds with the same spectrum (isospectral)?

\section{Heat Invariants}
\label{heat}

Let ${\cal M}$ be a smooth compact Riemannian manifold without boundary of 
dimension $m$. The Laplacian on this manifold has a spectrum $\{ \lambda_i \}$. 
Let us define the function (trace of the heat kernel):
\begin{equation}
Z(t) = \sum_i e^{-t \lambda_i}
\end{equation}
Taking the limit $t \rightarrow 0$ of the $Z(t)$ we get the 
Minakshisundaram-Peijel expansion \footnote{Also known in physics as the Schwinger-de Witt expansion (see \cite{v92} for a field theory perspective of the heat kernel). The name varies 
in the literature. Besides Schwinger-de Witt and Minakshisundaram-Peijel, it is sometimes 
also called the Minakshisundaram-Seeley expansion.}:
\begin{equation}
Z(t) = \frac{1}{(4 \pi t)^{m/2}} (a_0 + a_1 t + a_2 t^2 + ...)
\end{equation}
where the $a_i$ are some Riemannian invariant functions of the curvature 
tensor and  its covariant derivatives \footnote{See among 
others \cite{bgm71, glp99}.}.  If the manifold has a boundary there is a term proportional to $t^{1/2}$ in the expansion proportional to the volume of the boundary:
\begin{equation}
a_{1/2} = - \frac{\sqrt{\pi}}{2} Vol(\partial {\cal M}) .
\end{equation}

\vspace{0.5cm}

Let us consider the case without boundary. The first tree terms are:
\begin{itemize}
\item The first term is just the volume of the manifold:
\begin{equation}
a_0 = \int_{\cal M} dx \sqrt{g} ,
\end{equation}
This term can be easily obtained by computing the heat kernel as the partition function of a scalar field in a quadratic approximation. That is, this term can be obtained by taking the heat kernel as the partition function of a free scalar field on the manifold and performing the path integral by expanding the fields to quadratic order.

\item The second term is the integral of the Ricci scalar:
\begin{equation}
a_1 = \frac{1}{6} \int_{\cal M} dx \sqrt{g} R,
\end{equation}
\item The third term is 
\begin{equation}
a_2 = \frac{1}{360} \int_{\cal M} dx \sqrt{g} (5 R^2 - 2  R_{ij}R^{ij} 
+ 2 R_{ijkl}R^{ijkl}),
\end{equation}
\end{itemize}

\vspace{0.5cm}

From this expansion we can extract some information that only depends on the masses of the Kaluza-Klein modes:
\begin{itemize}
\item The dimension of the manifold can be read off from the pole at $t=0$. The divergence comes from the growing of the number of modes at high energy. The higher the number of dimensions, the stronger is the divergence.
\item From the $a_0$ term one can obtain the volume. Similarly, in a case of a manifold with boundary one can obtain the volume of the boundary from the $a_{1/2}$ term. 
\item From the $a_1$ term we obtain the integral of the Ricci scalar. In the 
particular case of dimension 2, this integral is related (by Gauss-Bonnet) to 
the Euler characteristic of the manifold: $a_1 = 2 \pi \chi({\cal M})/3$. Then 
in two dimensions the spectrum of the Laplacian encodes topological 
information -- the genus. Put it in another way, two isospectral manifolds 
in 2 dimensions have the same genus.
\item The $a_2$ term has no interpretation as a topological invariant in any 
dimension. The same is happening with the higher order terms. However, when wisely combined they can be used to get some topological information. That is the usual trick in some index theorems \footnote{See, for instance, the book of Gilkey \cite{gilkey84}.}.
\end{itemize}

In particular, two isospectral manifolds have the same dimension and volume. 
This asymptotic expansion does not contain all the topological information. 
For instance, a flat two dimensional torus and a flat Klein bottle have 
different spectrum but the asymptotic expansion will be the same if they have 
the same volume \footnote{All the $a_i$ vanish except the $a_0$ term due to 
the dependence on the curvature.}. In the asymptotic expansion we loose information that is encoded in terms of the form $e^{l_i/t}$, where $l_i$ are the lengths of closed geodesics. When, $t \rightarrow 0$ all these terms vanish like 'non-perturbative' effects. In section \ref{strings},
this relation between closed geodesics and the heat kernel will be used to discuss the masses of winding states in string theory. 

In general, there are some properties that can be obtained from the spectrum, such
as the dimension, volume, curvature, genus in dimension 2, etc. These are
the audible 
properties. There are other properties that are not audible, e.g., the first homotopy 
group \cite{v80}.

\section{Examples}

Now we are going to illustrate the behavior of the trace of the heat kernel and with two known examples. The first is a flat torus in any dimension. There the asymptotic behavior is particularly easy to obtain by using the Poisson resummation formula. The second example is the case of round spheres in any dimension where the asymptotics are a little bit more involved.

\label{examples}

\subsection{Flat Tori}

The easiest example is a flat $n$ dimensional torus. The spectrum is just:

\begin{equation}
\lambda_m = \sum_{ij} G^{ij} n_i n_j
\end{equation}
where the $n_i$ are integer numbers. The heat trace is:

\begin{equation} 
Z(t) = \sum_{n_i \in Z} e^{-t  \sum_{ij} G^{ij} n_i n_j}
\end{equation}
By using Poisson resummation and taking the limit $t \rightarrow 0$ we obtain the asymptotic expansion:

\begin{equation} 
Z(t) \sim \frac{\sqrt{Vol({\cal M})}}{(4 \pi t)^{n/2}}
\end{equation}

The only term that is surviving in this expansion is the first term. Higher order terms vanish due to the dependence on the curvature. In particular, we can deduce for the two dimensional torus that its Euler characteristic is zero.

When Poisson resummation is used we can see all the 'non-perturbative' \footnote{Non-perturbative in the sense that they cannot be obtained in the small $t$ expansion by perturbations around the zero.} terms $e^{-l_i/t}$  giving information about the closed geodesics (in string theory they are related to the winding modes). In the limit $t \rightarrow 0$ all these terms are exponentially suppressed.

\subsection{Spheres}

Let us consider a sphere of dimension $n$ with the usual round metric and radius $1$. The eigenvalues of the Laplacian are $\lambda_j = j (j +n -1)$. They appear with multiplicity: $deg_j = \left( \frac{n+j}{n} \right) - \left( \frac{n+j-2}{n} \right)$. The heat trace is just:

\begin{equation} 
Z(t) = \sum_j deg_j e^{-t j (j + n - 1)}
\end{equation}
The more familiar case of a round $S^2$ gives:
\begin{equation} 
Z(t) = \sum_j (2 j + 1) e^{-t j (j + 1)}
\end{equation}

The asymptotic expansion $t \rightarrow 0$ for the two dimensional sphere is \cite{ms67} \footnote{That can be computed by using Mellin transform, for instance.}:

\begin{equation} 
Z(t) = \frac{1}{t} \frac{e^{t/4}}{\sqrt{\pi t}} \int_0^1 
\frac{e^{-x/t}}{sin(\sqrt{x})} dx = (1/t + 1/3 + t/15 +...)
\end{equation}

In particular, one can deduce from here that the Euler characteristic is $\chi =2$.

\vspace{1cm}

For a three dimensional round sphere the asymptotic expansion gives \cite{ms67}:

\begin{equation} 
 Z(t) = \frac{Vol ({\cal M})}{(4 \pi t)^{3/2}} e^t = \frac{Vol ({\cal M})}{(4 \pi t)^{3/2}} (1 + t + t^2/2 +...)
\end{equation}

That is very easy to compute by writing:

\begin{equation} 
Z(t) = \sum_{j \geq 0} (j + 1)^2 e^{-t (j +1)^2 + t} = \frac{e^t}{2} \sum_{j \in Z} (j + 1)^2 e^{-t (j +1)^2} = -\frac{e^t}{2} \frac{d}{dt}\sum_{j \in Z} e^{-t j^2}
\end{equation}

and using the Poisson resummation formula:

\begin{equation} 
Z(t) = -\frac{e^t \sqrt{\pi}}{2} \frac{d}{dt}(t^{-1/2}\sum_{j \in Z} e^{- \frac{\pi^2 j^2}{t}}).
\end{equation}

That gives the correct asymptotic behavior by taking into account that the volume of a  round $S^3$ of radius 1 is $Vol ({\cal M}) = 2 \pi^2$.

\section{Higher Forms and Other Fields}
\label{forms}

In addition to functions, the Laplacian can also act on higher p-forms and their corresponding
spectra can give us further information about the compact manifold.
For instance, the size and the shape of a round sphere
in any dimensions is specified by 
the spectrum of the Laplacian on functions together with the spectrum of the Laplacian on one 
forms \cite{p70}. Another information than one can get immediately are the Betti numbers, because we know that the number of harmonic p-forms, $\Delta_g C = 0$, is just 
the Betti number $b_p$ of the manifold (Hodge-de Rham theorem).  So in 
principle, the zero modes alone, i.e. massless p-forms in four dimensions, 
can tell us the Betti numbers of the manifold. So, what is the information we get from higher forms? 

The asymptotic expansion can be carried out as in the case of functions, but 
now the coefficients are different \cite{gilkey84,p70,bgm71}:

\begin{equation}
Z_p(t) = \frac{1}{(4 \pi t)^{m/2}} (a_0 + a_1 t + a_2 t^2 + ...)
\end{equation}
where:

\begin{eqnarray}
a_0 & = & c(n,p)  Vol ({\cal M}) \nonumber \\
a_1 & = & \frac{c_0(n,p)}{6} \int_{\cal M} dx \sqrt{g} R \nonumber \\
a_2 & = & \frac{1}{360} \int_{\cal M} dx  (c_1(n,p) R^2 + c_2(n,p)  R_{ij}R^{ij} + c_3(n,p) R_{ijkl}R^{ijkl}) 
\end{eqnarray}

where the coefficients can be obtained from the function:

\begin{equation}
c(n,p)  = \left( 
\begin{array}{c} 
n \\ p 
\end{array}
\right) = \frac{n!}{(n-p)! p!}.
\end{equation}

The coefficients are:

\begin{eqnarray}
c_0(n,p) & = & c(n,p) - 6 c(n-2,p-1) \nonumber \\
c_1(n,p) & = & 5 c(n,p) - 60 c(n-2,p-1) + 180 c(n-4,p-2)\nonumber \\ 
c_2(n,p) & = & -2 c(n,p) +180 c(n-2,p-1) -720 c(n-4,p-2)\nonumber \\
c_3(n,p) & = & 2 c(n,p) - 30 c(n-2,p-1) + 180 c(n-4,p-2)
\end{eqnarray}

The non-trivial information on the manifold that is not present in the Laplacian on functions comes from the second term $a_2$ that gives a different linear combination of terms quadratic
in the curvature. 

By using Poincare duality one can relate the coefficients for p-forms to the coefficients for $(n-p)$-forms: $Z_p(t) = Z_{(n-p)}(t)$. So only half of the forms give us new information.

In particular the Laplacian acting on 1-forms gives rise to the following terms:

\begin{eqnarray}
a_1 & = & \frac{n-6}{6} \int_{\cal M} dx \sqrt{g} R \nonumber \\
a_2 & = & \frac{n}{360} \int_{\cal M} dx  (5 R^2 - 2  R_{ij}R^{ij} 
+ 2 R_{ijkl}R^{ijkl}) \nonumber \\
 & &  - \frac{1}{12} \int_{\cal M} dx \sqrt{g} (2 R^2 - 6  R_{ij}R^{ij} 
+ R_{ijkl}R^{ijkl})
\end{eqnarray}

Thus Laplacian on forms can be used to distinguish isospectral manifolds that are not locally isometric. If the manifolds are locally isometric, all the heat kernel asymptotics are the same as they depend only on the curvature.

\section{Isospectral Manifolds}
\label{isospectral}

M. Kac asked the following question in 1966 \cite{k66}: Can we hear the shape 
of a drum? If someone is playing a drum of a particular shape in another room, can 
we reproduce the shape of this drum by measuring the notes? This is a
 problem of finding the spectrum of a two-dimensional surface with Dirichlet 
boundary conditions. The answer is negative and was shown very recently in 1991 by 
C. Gordon, D. Webb and S. Wolpert \cite{gww92}. 
They explicitly constructed a pair of drums 
with different shape but the same spectrum.

 Here we will consider compact manifolds without boundary.  The first example 
of isospectral manifolds was found in 1964 by J. Milnor \cite{m64}: two 16 
dimensional flat tori \footnote{As we will see later these are the flat tori 
associated with the lattices of $E_8 \times E_8$ and $Spin(32)$ algebras. As we all know the partition function is the same for both theories. The distinction comes in the three point function,
i.e., the interaction.}. 
Later, other examples of flat tori were found in dimension 
four \cite{s90, cs92}. Isospectrality however is a rare phenomenon: almost all flat tori are uniquely determined by the spectrum and there are at most finitely many tori with the same spectrum \cite{w78}.

An example was found in 1980 \cite{v80} 
which illustrated 
that the first  homotopy group is not determined by the spectrum. 
The first examples of continuous isospectral 
metrics were found in 1984 by \cite{gw84}. 
Contrary to continuous isospectral metrics is the notion of 
spectral rigidity, i.e. there are 
no deformations of the metric that preserve the spectrum. For instance, all 
round spheres are spectrally rigid \cite{t80}.

There are a few Riemannian manifolds that are completely determined by 
their spectra: for instance, all the rounded
spheres $S^n$ with $n<6$ \cite{t73}. That can be generalized to all the 
rounded spheres 
by considering not only the spectrum on functions but also on one 
forms \cite{p70}.

One of the methods used to generate isospectral manifolds is the Sunada 
method \cite{s85}.  The basic idea is to start with a Riemannian manifold 
and to quotient it by two different discrete subgroups of the isometries, 
in such a way that the actions are free and there is a relation 
between the representations  \footnote{More specifically, if there is an unitary isomorphism between the representations of the groups on $L^2$-functions. Then it is easy to see that this construction gives isospectral manifolds: as the orbifold groups commute with the isometries, the Laplacian is invariant and the orbifold groups act on the eigenspaces of the Laplacian. The eigenfunctions 
should be invariant under the group.
If the dimension of the invariant states are the same, then the two manifolds are isospectral.}. 
Then the two orbifolds are isospectral. All 
the isospectral manifolds constructed in this way have a common Riemannian 
covering and they are always locally isometric. These manifolds can only be 
distinguished by global properties such as the first homotopy group if the two orbifold groups are not isomorphic. The Sunada method implies not only 
isospectrality on function but on p-forms \cite{tg89}.

Recently new isospectral manifolds have been constructed that are not
locally isometric. The first example dates back to 1991 which involves a pair of isospectral 
manifolds with boundaries \cite{s91}. In 1992 the first 
example of compact isospectral manifolds without boundary was constructed \cite{g93}. The 
method used in these construction is quite different from the one of Sunada 
and does not imply local isometry. This method was used to obtain continuous 
multi-parameter families of isospectral products of spheres (of dimension 
bigger than three) and tori (of dimension bigger than one) \cite{ggsww98}, 
and some simply connected closed isospectral manifolds \cite{s99}. However, the 
heat invariants for the Laplacian operator acting on one forms are different.

Recently isospectral non-locally isometric examples have been found in 
dimension four: $S^2 \times T^2$ and other interesting examples involving 
continuous isospectral families of metrics on Lie groups \cite{schueth}.  
In particular, the heat invariant for the Laplace operator on 1-forms changes. 
Isospectral metrics have been found for higher dimensional spheres and balls, 
Riemann surfaces of genus $g >3$, and etc.

\vspace{0.5cm}

We have seen, in analyzing the heat kernel expansion at $t \rightarrow 0$, that two isospectral manifolds have the same dimension (that can be read from the order of the pole at $t=0$), the same volume (that is the constant in the first term of the expansion), the integral of the Ricci scalar (the second term), and so on. In same cases, such as in two dimensions, one can automatically read off some topological data, e.g., the Euler character.

Furthermore, one can prove for two isospectral manifolds of dimension less than six that if one has constant sectional curvature \footnote{The sectional curvature is a curvature that is associated to two planes at each point. It can be defined in terms of the Riemann curvature as: $K_{ij} = R_{ijij}/(g_{ii} g_{jj} - g_{ij}^2)$. For instance in a two dimensional manifold is $K = R_{1212}/det(g) = R/2$.} , the other has constant sectional curvature as well \cite{t73,gilkey84}. 
In particular, if one is isometric to the standard sphere or a real projective space $RP^n$ so is the other.

It is easy to see from the heat kernel expansions that if two manifolds are isospectral for functions, 1 and 2-forms \cite{p70} that if one of the manifolds has constant scalar curvature 
or if it is Einstein
so is the other \footnote{That can be proved very easily by taking the $a_2$ terms of the expansions for the different forms and observing that the combinations of the curvature are linearly independent.}. For instance, we can ask ourselves, is there a
manifold isospectral to a K3 that is not another K3? If this is isospectrality in functions, one-forms and two forms we can use this theorem and the fact that the Betti numbers are the same to conclude that if there is such isospectral manifold it should be another K3.

We will see in section \ref{strings} same explicit examples of isospectral flat tori in dimension four and sixteen.

\section{Field Interactions}
\label{interactions}

Previously we have considered free fields on a compact manifold. Can 
interactions distinguish isospectral manifolds? As the simplest example, let us take a single 
scalar field with a cubic interaction term:

\begin{equation}
{\cal L} = \frac{1}{2} (\partial \phi)^2 + \frac{g}{3!} \phi^3.
\end{equation}
When reduced to four dimensions the field $\phi = \sum_i a_i \phi_i$ is 
decomposed into an infinite set of four dimensional scalar 
fields $a_i : R^{1,3} \rightarrow R$,  where the $\phi_i$ are the 
eigenfunctions of the Laplacian operator in the compact space. The 
interaction term in four dimensions looks like:

\begin{equation}
\frac{g}{3!}  \sum_{ijk} c_{ijk}  a_i a_j a_k
\end{equation}
where $c_{ijk} = \int_{\cal M} \phi_i \phi_j \phi_k$. The interaction with the massless fields $c_{0jk} = \delta_{jk}/\sqrt{Vol ({\cal M})}$ do not distinguish isospectral manifolds. But in general, the $c_{ijk}$ could make the distinction. 

\begin{figure}
\centering 
\epsfxsize=3in \hspace*{0in}\vspace*{.2in}
\epsffile{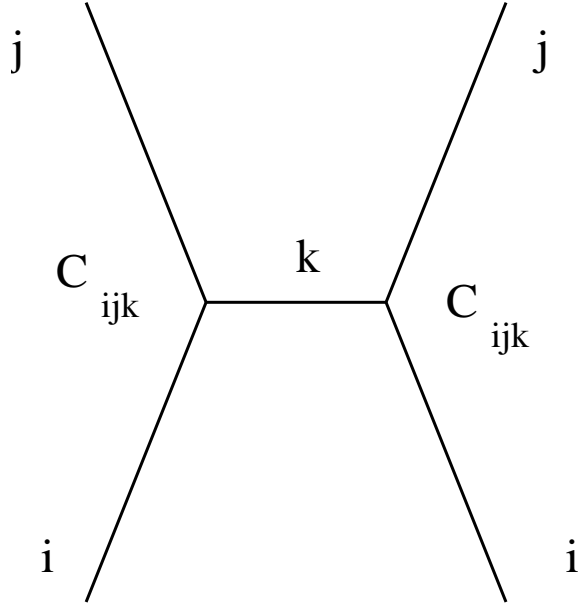}
\caption{\small 
Tree level scattering of Kaluza-Klein modes.}
\label{iijj}
\end{figure}

Notice that the massive particles do not enter in tree level processes of massless particles due to the $c_{0jk} = \delta_{jk}/\sqrt{Vol ({\cal M})}$. For this toy model, tree level involving massless fields processes are blind to higher modes and in particular they cannot distinguish isospectral 
manifolds \footnote{This is true for our simple model involving only a scalar field.
However, for more complicated cases involving gauge fields such as string theory on
$E_8 \times E_8$ and $SO(32)$, even the 
interactions between massless states can distinguish between 
isospectral manifolds, e.g., the coupling of three gauge bosons.}.
However the tree level interaction of massive modes depends on $c_{ijk}$. For instance, one can consider the annihilation of two $a_i$ modes to get two other modes $a_j$ (see figure \ref{iijj}):

\begin{equation}
{\cal M}_{ii \rightarrow jj} \sim g^2 \sum_{k} c_{iik} c_{jjk} \frac{1}{p^2 + \lambda_k}  
\end{equation}

%%%%%%%%%%%%%%%%%%%%%%%%
%%%%%%%%%%%%%%%%%%%%%%%%%

By using:

\begin{equation}
\frac{1}{p^2 + \lambda_j} = \int_0^{\infty} dt e^{-t (p^2 + \lambda_j)}
\end{equation}

one can express the amplitude in terms of a heat kernel:

\begin{equation}
{\cal M}_{ii \rightarrow jj} \sim g^2 \sum_{k} c_{iik} c_{jjk} \int_0^{\infty} dt e^{-t (p^2 + \lambda_k)} = g^2  \int_0^{\infty} dt e^{-t p^2} \sum_{k}  c_{iik} c_{jjk} e^{-t \lambda_k}
\end{equation}

\vspace{1cm}

\begin{figure}
\centering 
\epsfxsize=3in \hspace*{0in}\vspace*{.2in}
\epsffile{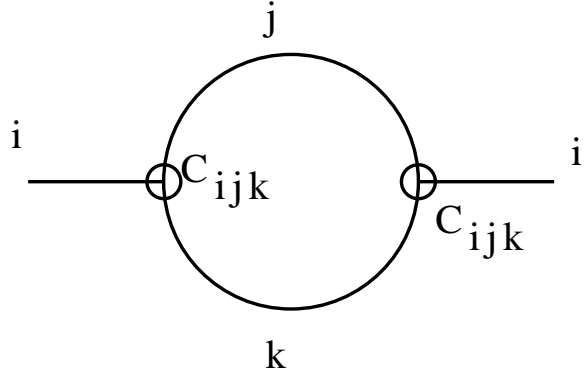}
\caption{\small 
Loop correction to the eigenvalues of the Laplacian.}
\label{loop}
\end{figure}

The higher modes enter, for instance, in the correction for the masses of the fields $a_i$ due to all the other modes circulating around a loop (see figure \ref{loop}):

\begin{equation}
{\cal M}_i \sim g^2 \sum_{jk} c_{ijk}^2 \int \frac{dp^4}{(p^2 + \lambda_j)(p^2 + \lambda_k)} 
\end{equation}

%%%%%%%%%%%%%%%%%%%%%%%%
%%%%%%%%%%%%%%%%%%%%%%%%%

One can see that the corrections can be written in terms of the function:

\begin{equation}
Z_i(t,t') =  \sum_{jk} c_{ijk}^2 e^{-t \lambda_j}e^{-t' \lambda_k}
\end{equation}

The difference between isospectral manifolds comes in the interaction coefficients $ c_{ijk}$ for massive modes. Let us take the corrections to the mass of the lowest state:

\begin{equation}
Z_0(t,t') =  \frac{1}{Vol ({\cal M})} \sum_{k}  e^{-(t+t') \lambda_k} = Z(t+t')
\end{equation}

That has the usual UV pole at short distances $t + t' \rightarrow 0$: 

\begin{equation}
Z_0(t,t') \sim  \frac{1}{(4 \pi (t+t'))^{n/2}} 
\end{equation}

\section{Strings on Isospectral Manifolds}  
\label{strings}

In general we cannot construct the spectrum of
string theory for two isospectral manifolds as we do not know the conformal theory description of them. However, as we have seen, there are some flat tori that are isospectral and 
non-isometric. The first known example is the 16 dimensional tori constructed 
by Milnor in 1964 \cite{m64} \footnote{These lattices are even and self-dual. 
As one suspects, the two lattices are the $E_8 \times E_8$ and $SO(32)$ 
lattices \cite{m64, bgm71} that appear in the heterotic strings\cite{heterotic}.
$E_8 \times E_8$ and $SO(32)/Z_2$ have the same partition function
but interactions (e.g. the algebra that they form) are different.
We can only tell the difference between the lattices at the level of interactions.}. We will analyze the  isospectral four dimensional tori of \cite{s90,cs92}. Then we will make some comments about more general backgrounds where we do not know the conformal theory description.

\subsection{Flat Tori}

Flat tori can be constructed by the quotient of flat space $R^n$ by a 
lattice. We will use isospectral lattices to describe a pair of lattices that give 
isospectral tori. In two dimensions the lattice is completely  determined 
by the theta-series.
In three dimensions the problem of finding two isospectral 
lattices is still open. The lowest dimension of two known flat  tori that are 
isospectral is in dimension four \cite{s90,cs92}. This is the case we will 
consider. Higher dimensional cases are known in 5, 6, 8, 12 and 16 dimensions.

Let us take the 4-parameter family of isospectral 4-dimensional tori 
constructed in \cite{cs92}. Let $e_i$, $i=1...4$, a set of orthogonal vectors 
satisfying the normalization:
\begin{equation}
e_i \cdot e_i = \frac{\alpha_i}{12} 
\end{equation}
where $\alpha_i > 0$. The lattices $\Lambda_{\pm}$ are generated by the 
vectors (written in the above basis):
\begin{eqnarray}
v_1^{\pm} & = & (\pm 3, -1, -1, -1) \nonumber \\
v_2^{\pm} & = & ( 1, \pm 3, 1, -1) \nonumber \\
v_3^{\pm} & = & ( 1, -1, \pm 3, 1) \nonumber \\
v_4^{\pm} & = & (1, 1, -1, \pm 3). 
\end{eqnarray}
The two lattices $\Lambda_{\pm}$ are isospectral.

The metrics on these lattices are:
\begin{equation}
G_{ij}^{\pm} = v_i^{\pm} \cdot v_j^{\pm} 
\end{equation}

So the Kaluza-Klein modes, $n_i$ plus the winding modes, $m^i$, have masses:
\begin{equation}
m^2 = G^{ij} n_i n_j + \frac{G_{ij}}{\alpha'^2} m^i m^j
\end{equation}

The dual lattice of $\Lambda^{\pm}$ with lengths $\alpha_i$ is 
$\Lambda^{\mp}$ with lengths $\alpha_i' = 1/\alpha_i$. 
One can easily see that they are also 
isospectral. So string theory cannot distinguish between this pair of 
isospectral manifolds.
In other words, if the lattices of momenta and windings 
are $(\Lambda^{+} (\alpha_i), \Lambda^{-} (1/\alpha_i))$, the isospectral 
lattices are $(\Lambda^{-} (\alpha_i), \Lambda^{+} (1/\alpha_i))$. The 
operation of mapping isospectral manifolds
commutes with T-duality and so the set of dualities on the torus is enlarged.

\subsection{Winding Modes}

In general, we do not have a conformal field theory description of two 
isospectral manifolds as the flat tori we have seen before, so the question 
for the general case of two isospectral metrics is still open.

Let us first address the following question: can winding modes distinguish 
isospectral manifolds?  Classically strings can wrap closed geodesic curves. 
The mass of the states associated with these strings is proportional to the 
lengths. We can order the closed geodesics $\beta_i$ by the length 
$l(\beta_i) \leq l(\beta_{i+1})$. The sequence $l(\beta_1), l(\beta_2), ...$ 
is called the length spectrum. Are the length and the eigenvalue equivalent? 
In our particular case we are interested in whether the spectrum of geodesic can 
be deduced from the eigenvalue spectrum.

In general, the two spectra are not equivalent \footnote{See, for 
instance, \cite{g86}.}.  But there are some cases, e.g., compact Riemann 
surfaces of genus $g > 1$ where the two spectra are equivalent.
For torus (genus 1), the momentum modes determine the windings in flat cases (just Poisson resummation).

But we are not interested in the exact equivalence between the two spectra 
but in the possible distinction of two isospectral manifolds from the 
classical winding states. For that, the theorem of Colin 
de Verdiere on closed geodesics \cite{cdv73} is very useful. Under the generic assumption 
that the set of closed geodesics are isolated and non-degenerated, it is 
proved that the eigenvalues of the Laplacian determine the spectrum of 
lengths of closed geodesics. 
So in a generic case, classical winding states 
cannot distinguish isospectral manifolds.

\section{What can we extract from a finite number of eigenvalues?}
\label{finite}

We have seen that even we cannot determine the complete geometry of a manifold 
by the spectrum one can obtain its dimension, volume, and etc.  We can then
wonder what kind of information about the manifold can we obtain with only a 
finite number of eigenvalues (the low-lying ones). There are several ways of understanding this problem. 

The first point of view is the one taken in the subject of
 finite spectral geometry \footnote{See for instance, the article of J.M. Lee 
in \cite{isg97} and \cite{l92,l93}.}.  We can try to estimate the dimension, volume or some constraints on the curvature. What are then  the properties of a manifold that we can estimate from a finite number of eigenvalues? How many 
eigenvalues do we need to estimate them with some given accuracy?

\begin{figure}
\centering 
\epsfxsize=5in \hspace*{0in}\vspace*{.2in}
\epsffile{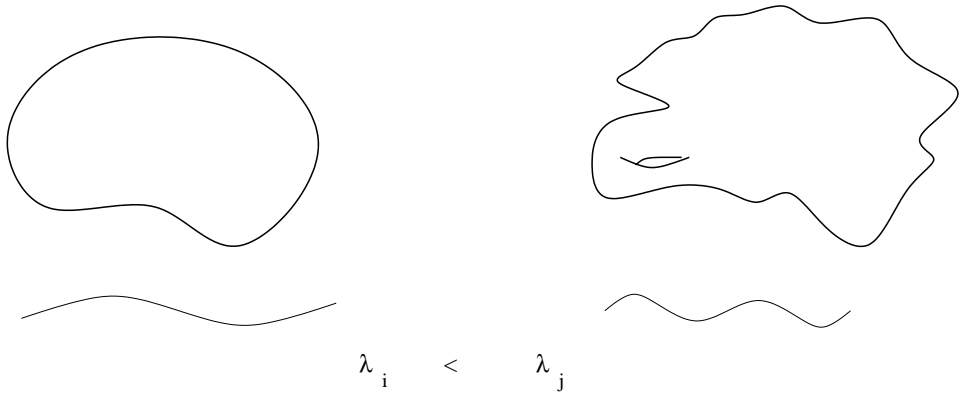}
\caption{\small 
Modes give information about the manifold at some scale $\lambda_i$. Higher modes $\lambda_j > \lambda_i$ give more detailed information about the manifold, i.e. information that cannot be resolved by lower modes.}
\label{lambda}
\end{figure}

The second viewpoint is more physical. In general, when we are 
measuring the properties of an object,
we use some scale and we define various properties of this object (like the volume, for instance) at this scale \footnote{A similar approach  has been taken in \cite{Dienes}, where the idea is
illustrated with flat tori.}. That scale is given in our case by the wavelength of the particle that is propagating in the manifold, which is of the order of the inverse of the mass of this particle. Very massive particles can give very detail information about the structure of the manifold  (see figure \ref{lambda}). Now imagine we want to measure the volume of our manifold. Using only the long wavelength modes we get an estimation of it at this scale. Taking into account higher order modes we can refine our measuring. From this point of view it is preferable to talk about the volume at some scale rather
than the true volume (whatever it is)  since perhaps the latter can never be defined. 

%%%%%%%%%%%%%%%%%%%%%%%%
%%%%%%%%%%%%%%%%%%%%%%%%%

The two viewpoints are very different but complementary. 
The viewpoint of finite spectral geometry assumes a fixed Riemannian manifold and by imposing bounds on some properties of the manifold one can obtain information about this manifold with a precision that depends on the bounds. 
The information on short scales is disregarded because of these bounds. The second viewpoint does not care about what is the exact structure at very small scales, like when we are measuring some gross properties of an object we do not care about the atomic structure of it, we just make a rough estimation at some scale. 
The second viewpoint also suggests that one needs more eigenvalues to determine
topological quantities such as the 
genus of a Riemann surface in contrast to the volume (as illustrated 
in figure \ref{lambda}). 

As an example take a round $S^2$ sphere of radius $R$. As we have seen in the examples, the eigenvalues of the Laplacian are $\lambda_j = j (j +1)/R^2$. The number of solutions with an eigenvalue less or equal to $\lambda_j$ are $N_j = (j +1)^2$ (counting multiplicities). Imagine that from other measurements we find out that there are only two dimensions. From the Weyl's formula we can approximate the volume of the manifold as follows:

\begin{equation} 
Vol ({\cal M}) \sim c_2 \frac{N_k}{\lambda_{N_k}} = 4 \pi R^2 (1 + \frac{1}{k})
\end{equation}
where $c_2 = 4 \pi$. So the masses of the fields give the following approximation to the volume of the manifold: $8 \pi R^2 $, $6 \pi R^2$, $16 \pi R^2 /3$,... Converging to the actual volume of the two sphere $4 \pi R^2$. In the more physical interpretation we can say that these values give the volume at different scales, but due to the convergence one can find a well defined limit $4 \pi$ and higher orders give very little information about the volume.

In what follows we will take the point of view of finite spectral geometry. In general we cannot deduce things like the number of extra dimensions, or the volume of the manifold without assuming something about the curvature of the manifold. That can be seen in a very simple example by J.M. Lee \cite{l93}: consider a Riemannian manifold ${\cal M}$ with some spectrum $\lambda_i$.  Now take a manifold ${\cal M} \times S^1_{\epsilon}$, where $S^1_{\epsilon}$ is a circle of radius $\epsilon$. The eigenvalues of the Laplacian on this new manifold are: $\lambda_i + n^2/\epsilon^2$. For $\epsilon$ small enough, the spectrum of the two manifolds agree in the first $N$ eigenvalues. This is an example of a family of finite isospectral manifolds. So in order to get some information about the manifold from a finite part of the spectrum some geometric assumptions should be made. As we will see, some of these assumptions can be that the 
curvature is bounded. 

The basic idea of \cite{ly80,l92,l93} is to take a formula involving a limit of the eigenvalues (like the Weyl formula or the heat trace expansion) and estimate the accuracy 
of measuring some properties of the manifold (e.g., the volume) with an error function.
Let us take, for instance, the Weyl formula and let us define the error function:

\begin{equation}
E(k) = \left| \frac{\lambda_k}{k^{2/n}} - \frac{c_n}{Vol ({\cal M})} \right|
\end{equation}

 Then one try to estimate $E(k)$, i.e., to show the existence of inequalities of the form $E(k) \leq f(k, c_i)$, where the $c_i$ are some quantities with a definite geometric meaning. For instance, the theorem of Li and Yau \cite{ly80} for convex Euclidean domains of fixed dimension (domains of $R^n$ where every two points are connected by a geodesic in the domain) says that there is a number $N$ such that higher modes $k > N$ satisfy $E(k) < \epsilon$. The number $N$ of eigenvalues needed to measure the volume of the domain depends only on $\epsilon$ and the eigenvalues $k < N$. 

One can repeat the argument for the heat trace and try to find some bound for the volume of a closed manifold. That is done in  \cite{l92,l93} by considering manifolds with some bound on the Ricci curvature, sectional curvatures, and so on.

In particular, we can get some bounds on the number of eigenvalues needed to 
obtain the dimension.  Let us suppose ${\cal M}$ to be a 
connected, compact Riemann manifold without boundary with sectional 
 curvature bounded from above and 
below $b > R_{sect} > \kappa > 0$, then there is a value $N$ of the low-lying 
eigenvalues from the spectrum of ${\cal M}$ such that the dimension is 
determined by them.

Another piece of interesting information is the spectral gap, i.e. the first eigenvalue 
$\lambda_1$\footnote{See, for instance, \cite{Bakry}.}. For instance, take the classical estimate of Lichnerowicz \cite{Lich}: if the Ricci curvature is bounded from bellow by $R \geq (n-1) \kappa > 0$, where $n$ is the dimension and  $\kappa$ is some positive real number, then the spectral gap is bigger than $\lambda_1 \geq n \kappa$. This inequality
becomes an equality only for the usual sphere \cite{Obata}. But this information is completely useless if $\kappa = 0$. 

More interesting is the result of Li \cite{Li} that says that if the Ricci scalar $R \geq 0$ then the spectral gap satisfies $\lambda_1 \geq \pi^2 /(2d^2)$ where $d$ is the diameter of the manifold. This limit has been improved  \cite{Zhong} to $\lambda_1 \geq \pi^2 /d^2$. That tells us that if we assume that the manifold has a metric with non-negative Ricci scalar curvature then we can get an estimation of the diameter from the first eigenvalue: $d \geq \pi / \sqrt{\lambda_1}$.

So, let us summarize the above discussions. In general, by taking the most general Riemannian manifolds one cannot say what is the dimension, the volume, and so on. However for reasonable manifolds (avoiding strong curvature, for instance) we can establish some bounds on the dimension and the volume of the manifold. It would be interesting to get other bounds on
the Euler characteristic for a Riemann surface from the asymptotic expansion of the heat trace. Also we have seen that interesting bounds on the diameter of the compact space can be obtained by the first eigenvalue, i.e. from the spectral gap, for positive Ricci scalar manifolds.

Another way of understanding this is if we want to obtain same details about the compactification structure from the long wavelengths modes (i.e., the lowest modes in the spectrum), we have to decouple the very energetic ones. Under some conditions these very massive modes do not contribute and we are able to give information about the shape of the manifold. The dimension and the volume become audible quantities.

\section{Conclusions}

In this paper we have explored many issues surrounding the question of how 
much information can one obtain from the spectrum (i.e. the masses) of the 
Kaluza-Klein modes. We have seen that one can reconstruct properties of the 
manifold such as the number of dimensions, the volume, and in some cases, 
like in two dimensional manifolds, topological information such as the genus. 
In reality, however, even if one day  we can detect these particles, 
we will have access to only a finite number of them. 
In this case we discussed how
one can reproduce some information about the manifold if certain 
assumptions are taken into account. 
In particular, if we assume that the manifold has a non-negative scalar curvature, then
the spectral gap gives us a lower bound on the size of the manifold.
In general, 
the modes at some mass $m = \sqrt{\lambda}$ give us information about the 
Riemannian manifold at scale $L \sim 1/\sqrt{\lambda}$. This wavelength 
puts a limit on the accuracy of our approximation. More detailed properties 
of the manifold that require a resolution smaller than $L$ remain
inaudible.

We have also analyzed some examples of manifolds with the same Laplacian 
spectrum, i.e. isospectral manifolds. These manifolds 'sound' the same and 
cannot be distinguished by the masses of all the massive modes. In some cases
where the conformal field theory for strings propagating in such isospectral 
backgrounds
can be explictly constructed, we have seen that string theory 
cannot 'hear' the difference between isospectral manifolds. 
The models relevant for string compactifications
that we have analyzed here are just isospectral flat tori.
General arguments tell us that the spectrum of the Laplacian can predict 
the spectrum of closed geodesics, i.e. the classical masses of the winding 
modes. 

We also pointed out that although the mass spectrum cannot distinguish 
between isospectral 
manifolds, generically interactions can hear the difference between them.
Furthermore, interactions also give rise to
corrections to the masses of the eigenmodes.

We have taken the 'very' toy model of a scalar field and p-forms on the 
compact manifold without boundary. Our analysis can be generalized to 
include fermions, gauge fields with non-trivial bundles, gravity, and etc. 
It would be interesting to extend our analysis to
manifolds with boundaries since in some cases 
the boundaries could have
the physical interpretation 
as branes \footnote{For example, the end-of-the-world branes
in the strongly coupled heterotic string theory \cite{HW}.}. 
Another interesting direction is to analyze the possible non-commutative deformations of a manifold. Following 
the work of Connes one can find non-commutative isospectral deformations of 
a manifold that in principle are spectrally rigid, e.g., a sphere.
For instance, it was shown in \cite{cl00} that any compact Riemannian spin manifold whose isometry group has rank $r \geq 2$ admits isospectral deformations to noncommutative geometries.

Many interesting questions remain: Can be hear the holonomy group of a manifold? Can we find two isospectral Calabi-Yau's? How can other fields distinguish isospectral manifolds? We hope to return
to some of these issues in the future.

\section*{Acknowledgments}

We thank Luis Alvarez-Gaum\'e, Roberto Emparan, Brian Greene, Dan Kabat, Heather Logan, 
Riccardo Rattazzi, Mark Trodden, Tom\'as Ort\'{\i}n and
Liantao Wang for discussions. RR thanks the University of Wisconsin for hospitality during the initial stage of this work.
The work of GS is supported in part by
funds from the University of Wisconsin.
GS also acknowledges the hospitality
of the Kavli Institute for Theoretical Physics 
during the final stage of this work.

\end{document}